\documentstyle[12pt]{article}
\def\STRIKE#1{\setbox0=\hbox{\ #1\ }\rlap{\hbox{\ #1}}\raise 3pt%
\hbox to\wd0{\hrulefill}}
\def\strikethru#1{\ifvmode\ifinner \STRIKE{#1}
                         \else    $\hbox{\STRIKE{#1}}$
                         \fi
                 \else   \ifmmode \hbox{\STRIKE{#1}}
                         \else    \STRIKE{#1}
                         \fi
                 \fi }
\begin{document}

\begin{center}
{\large
\bf Observation of contemporaneous optical radiation from 
a $\gamma$-ray burst}
\end{center}
\medskip

\begin{center}
C.~Akerlof$^\ast$, R.~Balsano$^\dagger$,
S.~Barthelmy$^{\ddagger\parallel}$, J.~Bloch$^\dagger$,
P.~Butterworth$^{\ddagger\#}$, \\ D.~Casperson$^\dagger$,
T.~Cline$^\ddagger$, S.~Fletcher$^\dagger$, F.~Frontera$^{\ast\ast}$,
G.~Gisler$^\dagger$, \\ J.~Heise$^{\dagger\dagger}$,
J.~Hills$^\dagger$, R.~Kehoe$^\ast$, B.~Lee$^\ast$,
S.~Marshall$^{\ddagger\ddagger}$,\\ T.~McKay$^\ast$,
R.~Miller$^\dagger$, L.~Piro$^{\parallel \parallel}$,
W.~Priedhorsky$^\dagger$,\\ J.~Szymanski$^\dagger$, J.~Wren$^\dagger$
\end{center}

\begin{tabular}{ll}
%a
$\ast$ &University of Michigan, Ann Arbor, MI 48109\\
%b
$\dagger$ &Los Alamos National Laboratory, Los Alamos, NM 87545 \\
%c
$\ddagger$ &NASA/Goddard Space Flight Center, Greenbelt, 
MD 20771, USA\\
%d
$\parallel$ &Universities Space Research Association, Seabrook, 
MD 20706, USA \\
%e
$\#$ &Raytheon Systems, Lanham, MD 20706 \\
%f
$\ast\ast$ &Universit\`{a} degli Studi di Ferrara, Ferrara, Italy \\
%g
$\dagger\dagger$ &Space Research Organization, Utrecht, The Netherlands \\
%h
$\ddagger\ddagger$ &Lawrence Livermore National Laboratory, 
Livermore, CA 94550 \\
%i
$\parallel\parallel$ &Instituto Astrofisica Spaziale, Rome, Italy \\
\end{tabular}
\bigskip
\hrule
\bigskip

\noindent{\it 
This paper has been submitted to and accepted for publication in
Nature, and is under embargo until it is published in that journal.
Other authors may reference this paper, but the contents should not be
reported in the popular media until the embargo is lifted.  If you
have any questions, contact Carl Akerlof
(akerlof@mich.physics.lsa.umich.edu).
}

\bigskip
\hrule
\bigskip

{\bf The origin of $\gamma$-ray bursts (GRBs) has been enigmatic since
their discovery$^1$.  The situation improved dramatically in 1997,
when the rapid availability of precise coordinates$^{2,3}$ for the
bursts allowed the detection of faint optical and radio afterglows -
optical spectra thus obtained have demonstrated conclusively that the
bursts occur at cosmological distances.  But, despite efforts by
several groups$^{4-7}$, optical detection has not hitherto been
achieved during the brief duration of a burst.  Here we report the
detection of bright optical emission from GRB990123 while the burst
was still in progress.  Our observations begin 22 seconds after the
onset of the burst and show an increase in brightness by a factor of
14 during the first 25 seconds; the brightness then declines by a
factor of 100, at which point (700 seconds after the burst onset) it
falls below our detection threshold.  The redshift of this burst,
$z\approx 1.6$ (refs 8, 9), implies a peak optical luminosity of
$5\times 10^{49}$ erg s$^{-1}$.  Optical emission from $\gamma$-ray
bursts has been generally thought to take place at the shock fronts
generated by interaction of the primary energy source with the
surrounding medium, where the $\gamma$-rays might also be produced.
The lack of a significant change in the $\gamma$-ray light curve when
the optical emission develops suggests that the $\gamma$-rays are not
produced at the shock front, but closer to the site of the original
explosion$^{10}$.}

The Robotic Optical Transient Search Experiment (ROTSE) is a programme
optimized to search for optical radiation contemporaneous with the
high-energy photons of a $\gamma$-ray burst.  The basis for such
observations is the BATSE detector on board the Compton Gamma-Ray
Observatory.  Via rapid processing of the telemetry data stream, the
GRB Coordinates Network$^{11}$ (GCN) can supply estimated coordinates
to distant observatories within a few seconds of the burst detection.
The typical error of these coordinates is $5^\circ$.  A successful
imaging system must match this field of view to observe the true burst
location with reasonable probability.

The detection reported here was performed with ROTSE-I, a two-by-two
array of 35 mm camera telephoto lenses (Canon {\it f}/1.8, 200 mm
focal length) coupled to large-format CCD (charge-coupled device)
imagers (Thomson 14 $\mu m$ 2,048 $\times$ 2,048 pixels).  All four
cameras are co-mounted on a single rapid-slewing platform capable of
pointing to any part of the sky within 3 seconds.  The cameras are
angled with respect to each other, so that the composite field of view
is $16^\circ \times 16^\circ$.  This entire assembly is bolted to the
roof of a communications enclosure that houses the computer control
system.  A motor-driven flip-away cover shields the detector from
precipitation and direct sunlight.  Weather sensors provide the vital
information to shut down observations when storms appear, augmented by
additional logic to protect the instrument in case of power loss or
computer failure.  The apparatus is installed at Los Alamos National
Laboratory in northern New Mexico.

Since March 1998, ROTSE-I has been active for $\sim$75\% of the total
available nights, with most of the outage due to poor weather.  During
this period, ROTSE-I has responded to a total of 53 triggers.  Of
these, 26 are associated with GRBs and 13 are associated with soft
$\gamma$-ray repeaters (SGRs).  The median response time from the
burst onset to start of the first exposure is 10 seconds.

During most of the night, ROTSE-I records a sequence of sky patrol
images, mapping the entire visible sky with two pairs of exposures
which reach a 5$\sigma$ V magnitude threshold sensitivity of $m_v =
15.$ These data, approximately 8 gigabytes, are archived each night
for later analysis.  A GCN-provided trigger message interrupts any
sequence in progress and initiates the slew to the estimated GRB
location.  A series of exposures with graduated times of 5, 75 and 200
seconds is then begun.  Early in this sequence, the platform is
`jogged' by $\pm8^\circ$ on each axis to obtain coverage of a four
times larger field of view.

At 1999 January 23 09:46:56.12 UTC, an energetic burst triggered the
BATSE detector.  This message reached Los Alamos 4 seconds later and
the first exposure began 6 seconds after this.  Unfortunately, a
software error prevented the data from being written to disk.  The
first analyzable image was taken 22 seconds after the onset of the
burst.  The $\gamma$-ray light curve for GRB990123 was marked by an
initial slow rise, so the BATSE trigger was based on relatively
limited statistics.  Thus the original GCN position estimate was
displaced by 8.9$^\circ$ from subsequent localization, but the large
ROTSE-I field of view was sufficient to contain the transient image.
At 3.8 hours after the burst, the BeppoSAX satellite provided an X-ray
localization$^{12}$ in which an optical afterglow was discovered by
Odewahn {\it et al.}$^{13}$ at Mt.~Palomar.  This BeppoSAX position
enabled rapid examination of a small region of the large ROTSE-I
field.  A bright and rapidly varying transient was found in the ROTSE
images at right ascension (RA) 15 h 25 min 30.2 s, declination (dec.)
44$^\circ$ 46'0'', in excellent agreement with the afterglow found by
Odewahn {\it et al.} (RA 15 h 25 min 30.53 s, dec. 44$^\circ$ 46'
0.5'').  Multiple absorption lines in the spectrum of the optical
afterglow indicate a redshift of $z>1.6.$ Dark-subtracted and
flattened ROTSE-I images of the GRB field are shown in Fig. 1.
Details of the light curve are provided in Table 1.

By the time of the first exposure, the optical brightness of the
transient had risen to $m_v = 11.7.$ The flux rose by a factor of 13.7
in the following 25 seconds and then began a rapid, apparently smooth,
decline.  This decline began precipitously, with a power-law slope of
$\sim$-2.5 and gradually slowed to a slope of $\sim$-1.5.  This
decline, 10 minutes after the burst, agrees well with the power-law
slope found hours later in early afterglow measurements$^{14}$.  These
observations cover the transition from internal burst emission to
external afterglow emission.  The composite light curve is shown in
Fig. 2.

A number of arguments establish the association of our optical
transient with the burst and the afterglow seen later.  First, the
statistical significance of the transient image exceeds 160$\sigma$ at
the peak.  Second, the temporal correlation of the light curve with
the GRB flux and the spatial correlations to the X-ray and afterglow
positions argue strongly for a common origin.  Third, the most recent
previous sky patrol image was taken 130 minutes before the burst and
no object is visible brighter than $m_v = 14.8.$ This is the most
stringent limit on an optical precursor obtained to date.  Searches
further back in time (55 images dating to 28 September 1998) also find
no signal.  Fourth, the `axis jogging' protocol places the transient
at different pixel locations within an image and even in different
cameras throughout the exposure series, eliminating the possibility of
a CCD defect or internal `ghost' masquerading as a signal.

The fluence of GRB990123 was exceptionally high (99.6 percentile of
BATSE triggers; M.~Briggs, personal communication), implying that such
bright optical transients may be rare.  Models of early optical
emission suggest that optical intensity scales with $\gamma$-ray
fluence$^{15-17}$.  If this is the case, ROTSE-I and similar
instruments are sensitive to 50\% of all GRBs.  This translates to
$\sim 12$ optically detected events per year.  Our continuing analysis
of less well-localized GRB data may therefore reveal similar
transients.  To date, this process has been hampered by the necessity
of identifying and discarding typically 100,000 objects within the
large field of view and optimizing a search strategy in the face of an
unknown early time structure.  The results we report here are at least
partially resolve the latter problem while increasing the incentive to
complete a difficult analysis task.  The ROTSE project is in the
process of completing two 0.45-m telescopes capable of reaching 4
magnitudes deeper than ROTSE-I for the same duration exposures.  If
gamma-ray emission in bursts is beamed but the optical emission is
more isotropic, there may be many optical transients unassociated with
detectable GRBs.  These instruments will conduct sensitive searches
for such events.  We expect that ROTSE will be important in the
exploration to come.

\begin{center}
Received 5 February; accepted 19 February 1999
\end{center}

%\begin{thebibliography}{99}
\begin{description}
\item{1.} Klebesadel, R.W., Strong, I.B., and Olson, R.A. Observations of
gamma-ray bursts of cosmic origin. {\em Astrophys. J.} {\bf 182,} 
L85-L88 (1973).

%none

\item{2.} Piro, L. {\em et al.} The first X-ray localization of a $\gamma$-ray
burst by BeppoSAX and its fast spectral evolution. {\em
Astron. Astrophys.}  {\bf 329,} 906-910 (1998), astro-ph/9707215.

\item{3.} Costa, E. {\em et al.} Discovery of an X-ray afterglow associated
with the $\gamma$-ray burst of 28 February 1997. {\em Nature} {\bf
387,} 783-785 (1997), astro-ph/9706065.

\item{4.} Krimm, H.A., Vanderspek, R.K., Ricker, G.R. Searches for optical
counterparts of BATSE gamma-ray bursts with the Explosive Transient
Camera.  {\em Astron. Astrophys. Suppl.} {\bf 120,} 251-254 (1996).

\item{5.}  Hudec, R. \& Sold\'{a}n, J. Ground-based optical CCD experiments
for GRB and optical transient detection.  {\em Astrophys. Space Sci.}
{\bf 231,} 311-314 (1995).

\item{6.} Lee, B. {\em et al.} Results from Gamma-Ray Optical 
Counterpart Search Experiment: a real time search for gamma-ray burst
optical counterparts. {\em Astrophys. J.} {\bf 482,} L125-L129 (1997),
astro-ph/9702168.

\item{7.} Park, H.S. {\em et al.} New constraints on simultaneous optical
emission from gamma-ray bursts measured by the Livermore Optical
Transient Imaging System experiment. {\em Astrophys. J.} {\bf 490,}
L21-L24 (1997), astro-ph/9708130.

\item{8.} Kelson, D.D. Illingworth, G.D., Franx, M. Magee, D., van 
Dokkum, P.G. {\em IAU Circ.} No.~7096 (1999).

\item{9.} Hjorth, J. {\em et al.~GCN Circ.} No.~219 (1999).

\item{10.}  Fenimore, E.E. Ramirez-Ruiz, E., Wu, B., GRB990123: Evidence 
that the gamma rays come from a central engine. Preprint
astro-ph/9902007 at $<$http://xxx.lanl.gov$>$ (1999).

\item{11.} Barthelmy, S. {\em et al.} in {\em Gamma-Ray~Bursts:~4th~
Huntsville Symp.} (eds. Meegan, C.A., Koskut, T.M. \& Preece, R.D.)
99-103 (AIP conf.~Proc.~428, Am. Inst. Phys., College Park, 1997).

\item{12.} Piro, L. {\em et al.~GCN~Circ.} No.~199 (1999).

\item{13.} Odewahn, S.C. {\em et al.~GCN~Circ.} No.~201 (1999).

\item{14.} Bloom, J.S. {\em et al.~GCN~Circ.} No.~208 (1999).

\item{15.} Katz, J.I. Low-frequency spectra of gamma-ray bursts. {\em
Astrophys. J.} {\bf 432,} L107-L109 (1994), astro-ph/9312034.

\item{16.} M\'{e}sz\'{a}ros, P. \& Rees, M.J. Optical and long-wavelength
afterglow from gamma-ray bursts. {\em Astrophys. J.} {\bf 476} 232-237
(1997), astro-ph/9606043.

\item{17.} Sari, R. \& Piran, T. The early afterglow. Preprint
astro-ph/9901105 at $<$http://xxx.lanl.gov$>$ (1999).

\item{18.} Bertin, E. \& Arnouts, S. SExtractor:  Software for source
extraction {\em Astron. Astrophys. Suppl.} {\bf 117,} 393-404 (1996).

\item{19.} H$\o$g, E. {\em et al.} The Tycho Reference Catalogue. 
{\em Astron.  Astrophys.} {\bf 335,} L65-L68 (1998).

\item{20.} Monet, D. {\em et al.} A Catalog of Astrometric 
Standards (US Naval Observatory, Washington DC, 1998).

\item{21.}  Zhu, J. \& Zhang, H. T. {\em GCN~Circ.} No.~204 (1999).

\item{22.} Bloom, J.S. {\em et al.~GCN~Circ.} No.~206 (1999).

\item{23.} Gal, R.R. {\em et al.~GCN~Circ.} No.~207 (1999).

\item{24.} Sokolov, V. {\em et al.~GCN~Circ.} No.~209 (1999).

\item{25.} Ofek, E. \& Leibowitz, E.M., {\em GCN~Circ.} No.~210 (1999).

\item{26.} Garnavich, P., Jha, S., Stanek, K., \& Garcia, M., 
{\em GCN~Circ.} No.~215 (1999).

\item{27.} Zhu, J. {\em et al.~GCN~Circ.} No.~217 (1999).

\item{28.} Bloom, J.S. {\em et al.~GCN~Circ.} No.~218 (1999).

\item{29.} Maury, A., Boer, M., \& Chaty, S. {\em GCN~Circ.} No.~220 (1999).

\item{30.} Zhu, J. {\em et al.~GCN~Circ.} No.~226 (1999).

\item{31.} Sagar, R., Pandey, A.K., Yadav, R.K.S., Nilakshi \& 
Mohan, V. {\em GCN~Circ.} No.~227 (1999).

\item{32.} Masetti, N. {\em et al.~GCN~Circ.} No.~233 (1999).

\item{33.} Bloom, J.S. {\em et al.~GCN~Circ.} No.~240 (1999).
\end{description}

\noindent
{\bf Acknowledgements} The ROTSE Collaboration thanks J.~Fishman and
the BATSE team for providing the data that enable the GCN
localizations which made this experiment possible; and we thank the
BeppoSAX team for rapid distribution of coordinates.  This work was
supported by NASA and the US DOE.  The Los Alamos National Laboratory
is operated by the University of California for the US Department of
Energy (DOE).  The work was performed in part under the auspices of
the US DOE by Lawrence Livermore National Laboratory.  BeppoSAX is a
programme of the Italian Space Agency (ASI) with participation of the
Dutch Space Agency (NIVR).

Correspondence and requests for materials should be addressed to
C.A. (e-mail: akerlof@mich.physics.lsa.umich.edu).
\pagebreak

\bigskip
\begin{table}

\caption{Exposure start times are listed in seconds, relative to
the nominal BATSE trigger time (1999 January 23.407594 UT). Exposure
durations are in seconds. Magnitudes are in the ``V equivalent''
system described in Fig.~1 legend.  Errors include both statistical
errors and systematic errors arising from zero point calibration. They
do not include errors due to variations in the unknown spectral slope
of the emission. Magnitude limits are 5 $\sigma$.  The final limit
results from co-adding the last four 200-s exposures and is quoted at
the mean time of those exposures. Camera entries record the camera in
which each observation was made.}
\begin{tabular}{c|ccc} \\
Exposure start &Exposure duration &Magnitude &Camera \\ \hline
-7922.08 &75 &$<$14.8 &C \\
22.18 &5 &11.70$\pm$0.07 &A \\
47.38 &5 &8.86$\pm$0.02 &A \\
72.67 &5 &9.97 $\pm$0.03 &A \\
157.12 &5 &11.86$\pm$0.13 &C \\
281.40 &75 &13.07$\pm$0.04 &A \\
446.67 &75 &13.81$\pm$0.07 &A \\
611.94 &75 &14.28$\pm$0.12  &A \\
2409.86 &200 &$<$15.6 &A \\
5132.98 &800 &$<$16.1 &A \\
\end{tabular}
\end{table}

~~
\pagebreak

\noindent
Figure 1.  Time series images of the optical burst. Each image is 24'
on a side, and represents $6\times 10^{-4}$ of the ROTSE field of
view.  The horizontal and vertical axes are the CCD pixel coordinates.
The sensitivity variations are due to exposure time; the top three
images are 5-s exposures, the bottom three are 75-s. The optical
transient (OT) is clearly detected in all images, and is indicated by
the arrow.  South is up, east is left.  Thermal effects are removed
from the images by subtracting an averaged dark exposure. Flat field
images are generated by median averaging about 100 sky patrol (see
text) images. Object catalogues are extracted from the images using
SExtractor.$^{18}$ Astrometric and photometric calibrations are
determined by comparison with the $\approx$1000 Tycho$^{19}$ stars
available in each image. Residuals for stars of magnitude 8.5-9.5 are
$<$ 1.2".  These images are obtained with unfiltered CCDs.  The optics
and CCD quantum efficiency limit our sensitivity to a wavelength range
between 400 and 1100 nm. Because this wide band is non-standard, we
estimate a ``V equivalent" magnitude by the following calibration
scheme.  For each Tycho star, a ``predicted ROTSE magnitude" is
compared to the 2.5 pixel aperture fluxes measured for these objects
to obtain a global zero point for each ROTSE-I image.  For the Tycho
stars, the agreement between our predicted magnitude and the measured
magnitude is $\pm$0.15.  These errors are dominated by colour
variation.  The zero points are determined to $\pm$0.02.  With large
pixels, we must understand the effects of crowding. (This is
especially true as we follow the transient to ever fainter
magnitudes.)  To check the effect of such crowding, we have compared
the burst location to the locations of known objects from the USNO A
V2.0 catalouge.$^{20}$ The nearest object, 34" away, is a star with
R-band magnitude $R = 19.2.$ More important is an R = 14.4 star, 42"
away. This object affects the measured magnitude of the OT only in our
final detection.  It can be seen in the final image to the lower right
of the OT.  A correction of +0.15 is applied to compensate for its
presence.  Magnitudes for the OT associated with GRB 990123, measured
as described, are listed in Table 1.  Further information about the
ROTSE-I observations is available at http://www.umich.edu/$\sim$rotse.

\pagebreak

\noindent
Figure 2.  A combined optical light curve. Afterglow data points are
drawn from the GCN archive.$^{21-33}$ The early decay of the ROTSE-I
light curve is not well fit by a single power law. The final ROTSE
limit is obtained by co-adding the final four 200-s images.  The inset
shows the first three ROTSE optical fluxes compared to the BATSE
$\gamma$-ray light curve in the 100-300 KeV energy band.  The ROTSE-I
fluxes are in arbitrary units.  Horizontal error bars indicate periods
of active observation.  We note that there is no information about the
optical light curve outside these intervals.  Vertical error bars
represent flux uncertainties.  Further information about GCN is
available at http://gcn.gsfc.nasa.gov/gcn.

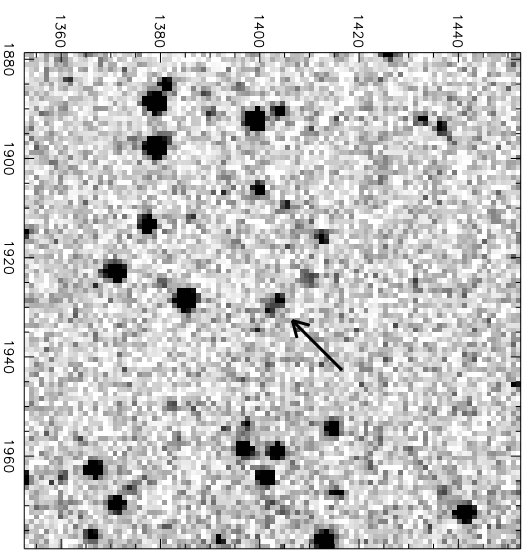
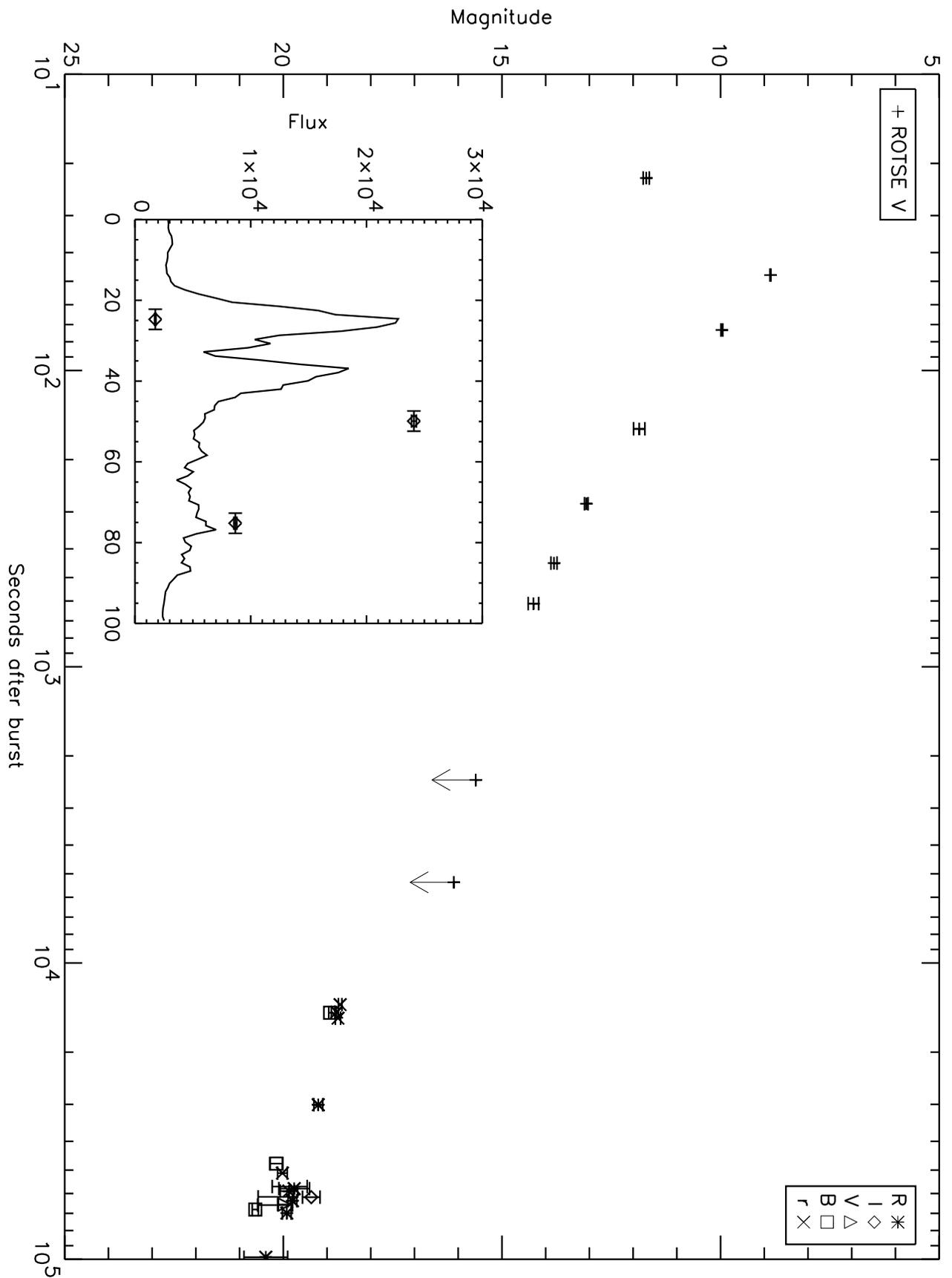
\end{document}